\documentclass[runningheads]{llncs}
\usepackage[T1]{fontenc}
\usepackage{graphicx}
\usepackage{soul}  
\usepackage{xcolor}
\usepackage{xspace}
\usepackage{booktabs}
\usepackage{colortbl}
\usepackage{amsmath, amssymb}

\begin{document}
\title{ Context-AI Tunes: Context-Aware AI-Generated Music for Stress Reduction}
\titlerunning{Context-AI Tunes}
%
\author{Xiaoyan Wei\orcidID{0000-0001-5535-4197} \and
Zebang Zhang\orcidID{0009-0009-6418-8720} \and
Zijian Yue\orcidID{0009-0009-2538-5836}
\and
Hsiang-Ting Chen\orcidID{0000-0003-0873-2698}}
\authorrunning{Wei, Zhang et al.}
%
\institute{The University of Adelaide, Adelaide 5007, AU \\
\email{\{xiaoyan.wei, tim.chen\}@adelaide.edu.au\\
\{zebang.zhang, zijian.yue\}@student.adelaide.edu.au}}

\maketitle              
\begin{abstract}
Music plays a critical role in emotional regulation and stress relief; however, individuals often need different types of music tailored to their unique stress levels or surrounding environment. Choosing the right music can be challenging due to the overwhelming number of options and the time-consuming trial-and-error process. 
To address this, we propose Context-AI Tune (CAT), a system that generates personalized music based on environmental inputs and the user’s self-assessed stress level.
A 2×2 within-subject experiment (N=26) was conducted with two independent variables: AI (AI, NoAI) and Environment (Busy Hub, Quiet Library). 
CAT’s effectiveness in reducing stress was evaluated using the Visual Analog Scale for Stress (VAS-S). 
Results show that CAT is more effective than manually chosen music in reducing stress by adapting to user context.

\end{abstract}

\section{Introduction}

Music plays an important role in alleviating stress and reducing anxiety, and people often turn to music in various settings to help manage their stress \cite{Watson2017,lee2013does}. However, most relaxing music available today is pre-set and does not directly relate to the environment or the stress level of the user. For instance, in a stressful work environment where one feels tired and anxious, selecting a piece of relaxing music on YouTube might offer brief relief, but it does not adapt to the listener’s current stress state or surroundings. Consequently, its effectiveness may decrease in different contexts.

Recent advancements in generative AI models such as large language models (LLMs) and generative music model have opened up new opportunities for generating adaptive, personalized content in real time~\cite{cao2023AIGC}. 
By using their ability to process various inputs—including user feedback and environmental context—these generative AI models can be utilized to create context-aware music tailored to an individual’s surroundings and stress level. These models enable systems to analyze contextual factors and user inputs to generate prompts for personalized music compositions, bridging the gap between static relaxing music and dynamic, user-centric interventions~\cite{hou2022ai}. With AI playing an increasingly prominent role in music generation and personalization, we propose the following research question: 
\begin{quote}
\textit{Does AI-generated music tailored to the user’s environment and stress level outperform pre-recorded relaxing music in effectiveness?}
\end{quote}
To answer the question, we introduce Context-AI Tune (CAT), a system that generates personalized relaxing music by integrating user inputs with environmental context (Figure~\ref{Main Interface}). 
The system uses a camera to capture the surrounding environment and extracts descriptive keywords via a visual language model (e.g., ChatGPT). Users select and combine these keywords, input their perceived stress level, and the Suno API generates adaptive music tracks accordingly~\cite{Suno}.
This approach enables CAT to deliver music tailored to both the user’s stress level and physical surroundings.

Through comparative experiments, we evaluated the effectiveness of CAT in reducing stress compared to traditional hand-picked relaxing music. 
A 2×2 within-subject experiment (N=26) was conducted with two independent variables: AI (AI, NoAI) and Neighbor (Busy-Hub, Quiet-Library). 
The study took place in two distinct environments: a busy and noisy student hub center and a quieter, more focused library. 
Participants completed stress-inducing math tasks using a Python-based program featuring timed problem solving, auditory feedback, and leaderboards to enhance competitiveness. 
Stress levels were measured using the Visual Analog Scale for Stress (VAS-S)\cite{Lesage_Berjot_Deschamps_2012}. 
Results indicated that CAT was more effective in relieving stress across both environments.

The contributions of this paper are as follows: 
\begin{itemize}
    \item We propose CAT, a system for generating personalized relaxing music based on environmental data and user stress levels.
    \item We conducted a within-subject experiment (N=26) using VAS-S to evaluate the impact of AI-generated music on stress reduction.
    \item Our results provide evidence that AI-generated music, taking both environmental context and user stress levels into account, offers greater potential for stress relief compared to traditional relaxing music.
\end{itemize}
\section{Related Work}
\subsection{Music Therapy in HCI Research}
Research in HCI has explored various applications of music and musical tools within the context of music therapy \cite{Baglione_Clemens_Maestre_Min_Dahl_Shih_2021}, including accessibility, education, and collaborative experiences. Recent studies have employed mobile sensors to enhance user interaction by providing auditory, visual, and tactile feedback, as well as augmenting traditional musical instruments to address diverse user needs.

From the early stages, music therapists integrated computer-based technologies to support music therapy interventions. For instance, the CAMTAS system organized audio and video data collected during therapy sessions, enabling therapists to track physical activities and analyze interactions quantitatively \cite{Verity2003ACA}. Similarly, tools such as the Music Therapy Toolbox analyzed MIDI recordings \cite{erkkila2007music}, while the Music-therapy Analyzing Partitura (MAP) facilitated qualitative analysis of therapy events \cite{gilboa2007testing}. As HCI research advanced, more user-centered approaches emerged in music therapy, focusing on user engagement \cite{kirk2016motivating} and interactivity \cite{hamidi2019sensebox,ragone2020designing} to enhance the overall experience. In addition, augmented reality \cite{correa2009computer,lobo2019chimelight}, virtual reality \cite{perez2022immersive,zhu2021stress,zhu2024stressBCI}, and tangible user interfaces \cite{hamidi2019sensebox,ragone2020designing,lobo2019chimelight,zhu2021drone} have increasingly been applied in this area.

Although these technologies have enriched music therapy with advanced features and user-centered principles, most systems remain limited in their adaptability to users’ real-time environments and stress states. Existing tools often rely on predefined settings or static interactions \cite{erkkila2007music,gilboa2007testing,hamidi2019sensebox}, lacking the dynamic capability to adjust based on the user’s context. To address this gap, our work utilized AI and multimodal inputs to generate personalized relaxing music—aiming to enhance stress relief by aligning music with both environmental cues and users’ stress levels.

\subsection{Generative AI in Music production}

Generative AI technologies, such as music recommendation \cite{schedl2015music}, music generation \cite{herremans2017functional}, and pitch modification \cite{engel2020ddsp}, have gained attention in the HCI community for their potential in various music-related applications. These technologies can be categorized into general melody generation \cite{hsiao2021compound,roberts2018hierarchical}, emotion-driven music generation \cite{grekow2021monophonic,hung2021emopia}, melody harmonization \cite{huang2019counterpoint}, and music genre or pitch transfer \cite{brunner2018midi,engel2020ddsp}. While advancements in this field are promising, their application in music therapy remains limited. Existing studies often focus on static algorithms, such as music recommendation \cite{biswal2021can,yuan2021application}, or training generative models to produce music for therapy \cite{hou2022ai,yu2023musicagent,williams2020use}, with minimal emphasis on the dynamic human-AI interaction needed for adaptive therapeutic contexts.

Examples of AI systems designed for therapy include DeepThInk, an AI-infused art creation system developed with art therapists to lower expertise barriers and enhance creativity \cite{du2024deepthink}. AI-powered interactive systems have also been introduced to support collaborative decision-making during rehabilitation assessments \cite{lee2020co,lee2021human} and to facilitate empathetic dialogues in mental health support \cite{sharma2023human}. These applications demonstrate the potential of AI to enrich therapeutic interventions through collaboration and personalization.

In the field of music therapy, there is growing interest in human-AI collaboration for creative activities such as live performances \cite{bian2023momusic,hanson2020sophiapop}, composition \cite{frid2020music,louie2020novice}, and production \cite{nicholls2018collaborative}. These studies highlight the capacity of generative AI to enhance both creative processes and therapeutic outcomes. However, its use in interactive music therapy—particularly in adapting music dynamically to users’ environments and stress levels—remains underexplored. This study aims to fill this gap by introducing a system that integrates generative AI with contextual and stress-related inputs to provide personalized relaxing music.

\section{System Design}

\begin{figure} [ht]
    \centering
    \includegraphics[width=\linewidth]{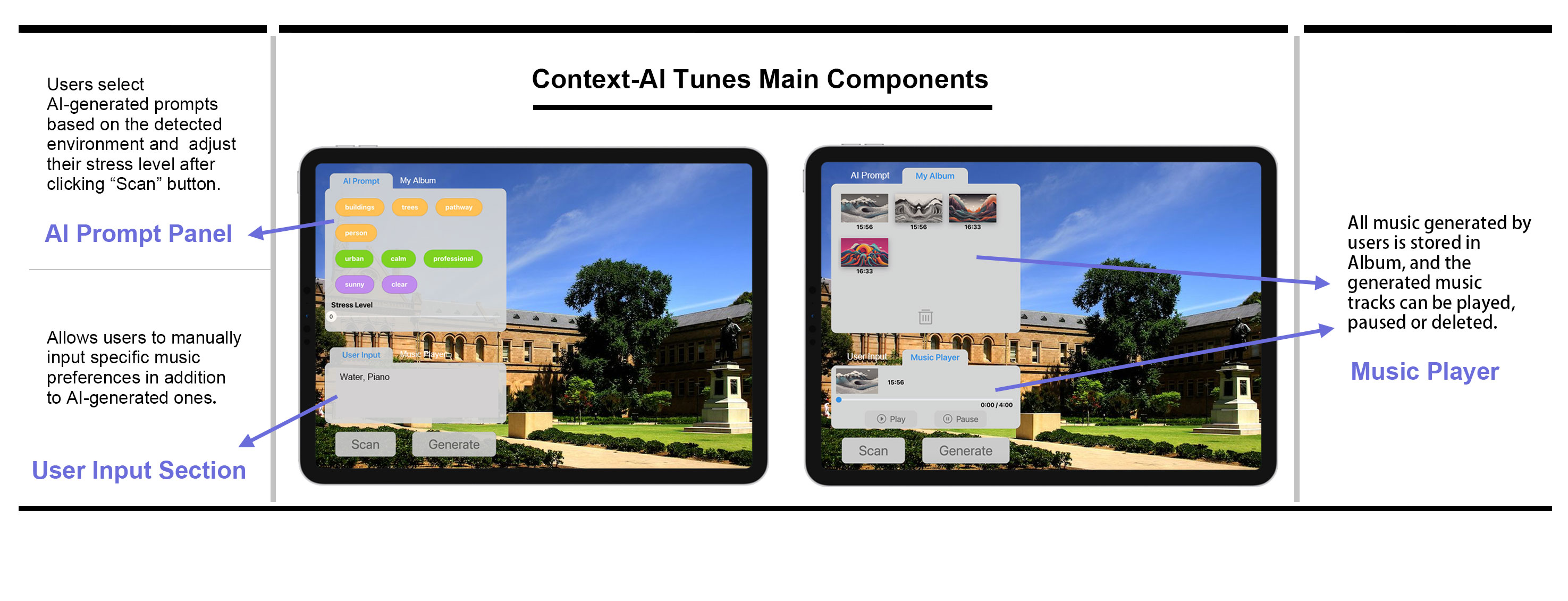}
    \caption{Main components of CAT}
   \label{Main Interface}
\end{figure}

The Context-AI Tune (CAT) system integrates a visual language model to generate personalized music tailored to users’ environments and stress levels. It comprises three main components: the AI Prompt Panel, the User Input Section, and the Music Player (Fig. \ref{Main Interface}). Below, we describe the functionality of each component through a typical user interaction scenario.

\begin{itemize}
    \item AI Prompt Panel: Users begin by clicking the “Scan” button, which activates the system’s camera to capture the surrounding environment. The system processes the captured image via ChatGPT to generate context-based prompts (e.g., “trees,” “building,” “pathway”) based on detected features. These prompts appear in the AI Prompt Panel for the user to review and select.

    \item User Input Section: After selecting a prompt, users can adjust their stress level through a slider and add descriptive keywords like "piano" or "water." These inputs allow users to specify preferences and personalize the generated music tracks. The interface is designed to make the input process straightforward and accessible.

    \item Music Player: Once the user inputs are complete, clicking the "Generate" button produces two music tracks based on the selected prompts, stress level, and keywords, using the Suno API. The generated tracks are saved in "My Album," where users can play, pause, or delete them using the music player.
\end{itemize}

For instance, a user working in a quiet library might select the “trees” prompt, set a medium stress level, and add keywords like “piano” and “calm.” The system would generate music tracks based on the environment and these inputs, offering a tailored auditory experience for stress relief.

This three-component structure illustrates how environmental data and user inputs can be effectively combined to create personalized music aimed at stress management.

\subsection{Implementation}

\begin{figure} [ht]
    \centering
    \includegraphics[width=\linewidth]{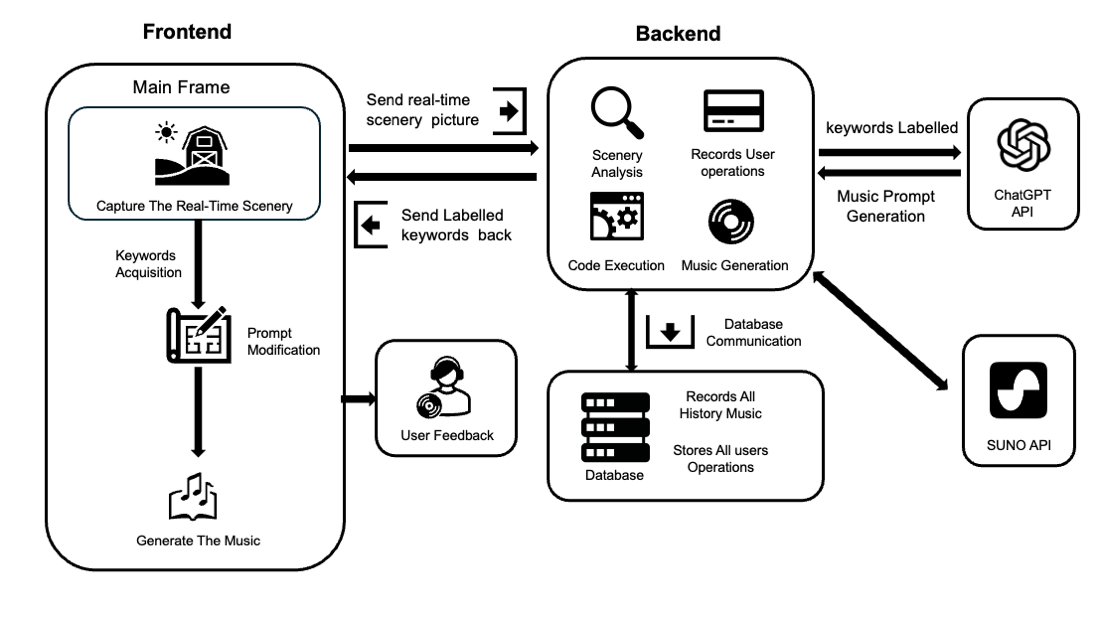}
    \caption{System Workflow}
   \label{Workflow}
\end{figure}

The CAT system features a dual-component architecture consisting of a Vue 3-based frontend and a Python Flask backend. The frontend allows users to capture real-time images using a canvas element, modify prompts, and interact with the music player. Images are encoded in Base64 format and sent to the backend for analysis. The frontend also supports live audio streams and MP3 playback, calculating progress to ensure proper functionality. User interactions, including prompt selection and music playback, are logged for experimental analysis.

The backend processes image data through the ChatGPT API to generate descriptive keywords, refining them into music prompts, and sends these prompts to the Suno API for music generation. An SQLite database stores user interactions and operational data, which are then exported as CSV files for further examination. This architecture ensures personalized music generation based on user inputs and environmental context.

\subsubsection{Environmental Analysis}
Captured images are passed to the ChatGPT API, which identifies environmental elements and produces keywords such as “trees,” “building,” or “pathway.” These keywords appear on the frontend for user review and modification. Once finalized, the backend compiles the keywords into a structured prompt for music generation, ensuring that the resulting tracks reflect both the user’s environment and stated preferences.

\subsubsection{Music Generation}
Music tracks are generated through the Suno API, which processes the prompts produced by the backend. Depending on the use case, these tracks can be delivered either as live streams or MP3 files. The backend handles the format conversion and streams the content to the frontend for playback.

\subsubsection{Data Logging and Analysis}
All user interactions—such as prompt selection, stress level adjustments, and playback actions—are recorded in the backend. These logs are maintained in a structured format and exported as CSV files for use in experimental evaluations and ongoing system improvements
\section{User Study}
We conducted a within-subject 2×2 experimental study with two independent variables: AI (AI, NoAI) and Environment (Busy Hub, Quiet Library) to investigate whether music generated via CAT is more effective in reducing stress compared to traditional relaxing music. 

\subsection{Participants}
We recruited 26 participants (10 male, 16 female) aged between 18 and 35 from our academic institution. 
Participants were compensated with a \$20 AUD gift card. The study was approved by the institute's Human Research Ethics Committee. 

\subsection{Experimental Setup}
The study was conducted in two distinct environments: a busy student hub(B-Hub) and a quieter library(Q-Lib). Each participant was provided with a laptop for completing tasks, an iPad for interacting with the CAT system, and a pair of noise-canceling headphones for listening to the generated music. Additionally, subjective stress levels were assessed using the VAS-S \cite{cella1986reliability} at multiple stages.

\begin{figure} [ht]
    \centering
    \includegraphics[width=\linewidth]{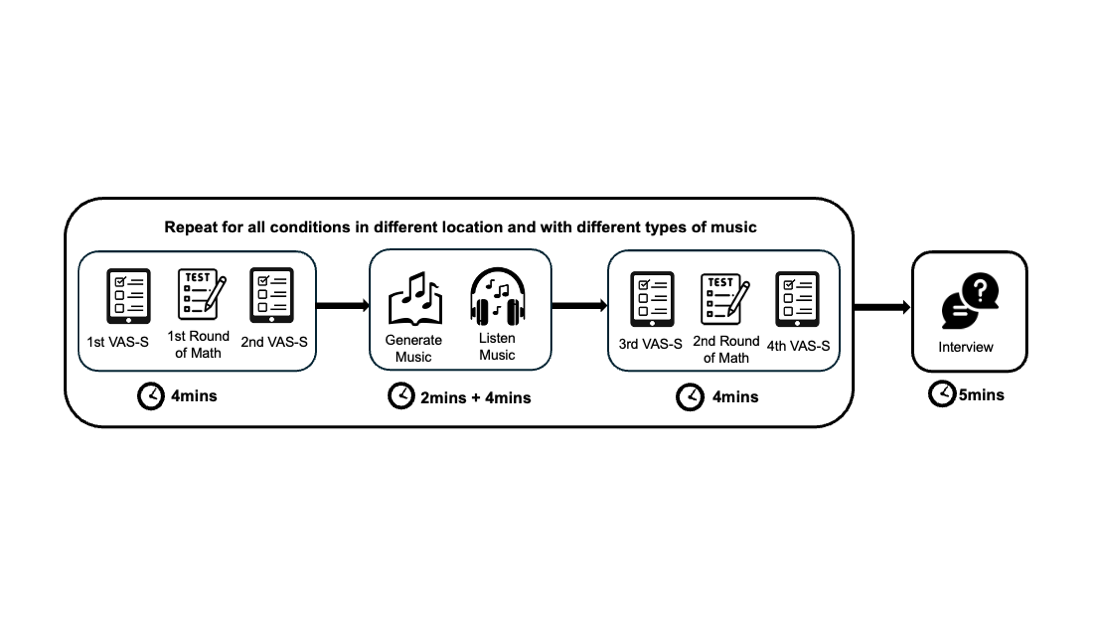}
    \caption{Experimental session flow, repeated four times for each participant across two different days. The process includes VAS-S questionnaire, math tasks, and a post-hoc interview to assess stress and engagement levels}
   \label{Study Process}
\end{figure}

To minimize potential learning effects from repeated exposure to the math tasks, the order of the conditions was randomized for each participant. This randomization ensured that the sequence of variables did not introduce bias into the experimental results, allowing for a more reliable evaluation of the hypothesis. 
The experimental session consisted of a series of tasks designed to induce and measure stress (Fig. \ref{Study Process}). Each session included the following stages:

Fig.~\ref{Study Process} illustrates the tasks involved in a complete session:

\begin{itemize}
    \item At the beginning of the session, participants completed the first round of VAS-S.
    \item Participants performed a math computation task consisting of 40 two-digit addition and subtraction problems with an 8-second time limit per problem. A countdown timer and auditory feedback were used to increase task-related tension.
    \item Participants recorded their second round of VAS-S.
    \item Participants either listened to personalized music generated by the CAT system or to one of the most popular relaxing music tracks selected from YouTube’s top-rated music channels. Each listening session lasted 4 minutes.
    \item Participants completed the third round of VAS-S.
    \item Participants performed another math task, followed by the fourth round of VAS-S.
\end{itemize}

To simulate a stressful environment, we developed a Python-based math computation task where participants solved 40 timed addition and subtraction problems involving two-digit numbers and decimals in each session. A leaderboard and auditory feedback were introduced to enhance motivation and task-related stress. Participants rated their stress levels at four stages during each session, and this sequence was repeated across all conditions for each participant.

\subsection{Interview}
After completing the experiment, participants took part in a 5-minute semi-structured interview focusing on:
\begin{itemize}
    \item Experience with the CAT system: Participants were asked about the system’s ability to reduce stress and the suitability of the generated music.
    
    \item Environment and Music Type: Feedback was gathered on how the environment and music type influenced their stress levels.
    
    \item Suggestions for Improvement: Participants shared insights on potential system enhancements and broader applications.
\end{itemize}
\section{Result}
Data from all 26 participants (10 male, 16 female) were analyzed. Each participant successfully completed all four experimental conditions, with no dropouts or incomplete data reported. The analysis focused on subjective stress levels measured by the VAS-S. Repeated-measures ANOVA was used to evaluate the main effects of music type (AI vs. NoAI) and environment (B-Hub vs. Q-Lib), as well as their interaction\cite{wobbrock2011aligned}. Pairwise comparisons were conducted using Wilcoxon Signed-Rank Test\cite{woolson2005wilcoxon} to identify specific differences between conditions\cite{elkin2021aligned}. The results provide insights into the effectiveness of CAT-generated music in reducing stress across different environments, which are detailed in the following sections.

\subsection{Result of VAS-S Score}

\begin{figure} [ht]
    \centering
    \includegraphics[width=\linewidth]{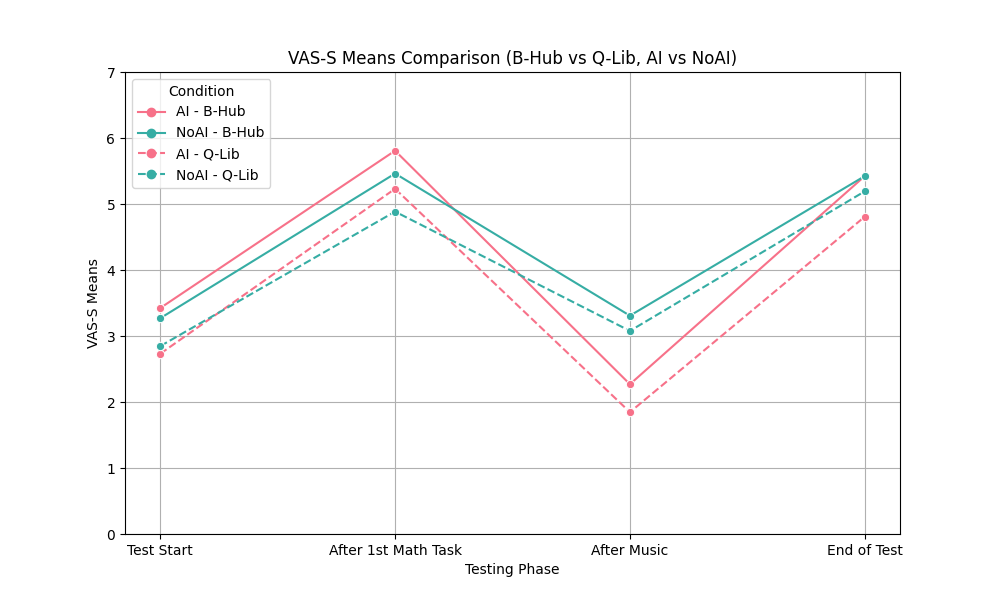}
    \caption{VAS-S Means by Condition (B-Hub vs Q-Lib, AI vs NoAI)}
   \label{VASlines}
\end{figure}

\begin{figure} [ht]
    \centering
    \includegraphics[width=\linewidth]{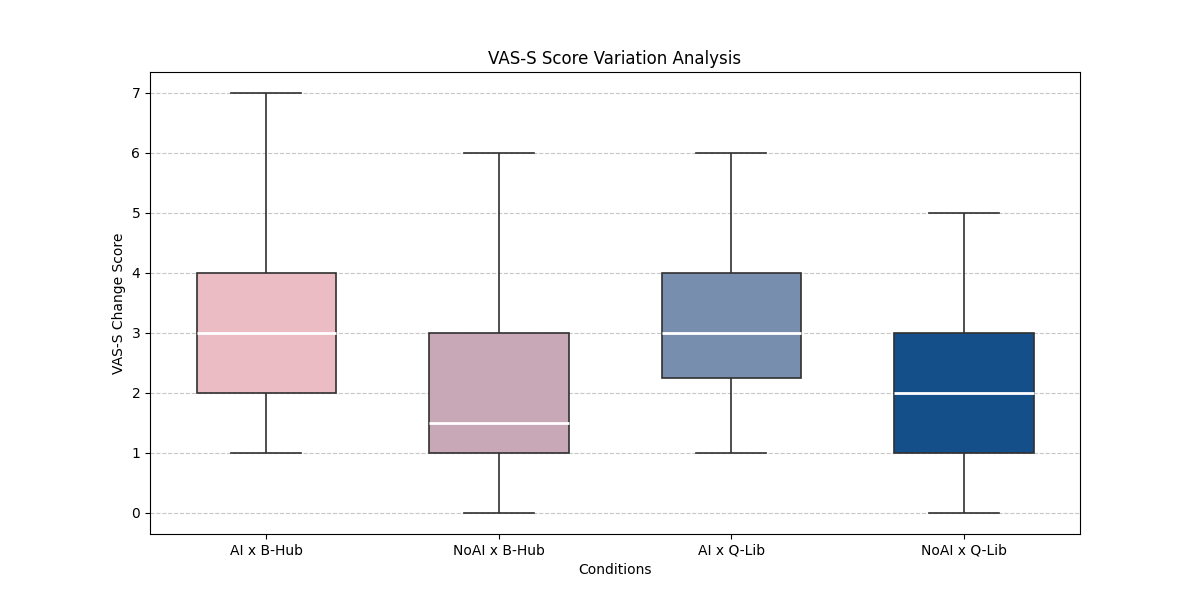}
    \caption{VAS-S Score Variation Analysis in Four Conditions }
   \label{VASChange}
\end{figure}

As shown in Fig. \ref{VASlines}, participants’ VAS-S scores were recorded at four key phases(Fig. \ref{Study Process}) (Test Start=1st VAS-S, After 1st Math Task=2nd VAS-S, After Music=3rd VAS-S, and End of Test=4th VAS-S) under four different conditions: AI-B-Hub, NoAI-B-Hub, AI-Q-Lib, and NoAI-Q-Lib.  At the beginning of the experiment (Test Start), most participants reported relatively moderate stress levels (around 2.5 to 3.5 on average), with only small differences across the two environments (B-Hub vs. Q-Lib) and the two music types (AI vs. NoAI). Once participants completed the first math task, all conditions exhibited a substantial increase in perceived stress from 2 to 5.5, reflecting the task’s difficulty and the influence of environmental distractions.

Following the music intervention (After Music), there was a marked decrease in VAS-S scores across all conditions, though the AI-generated music demonstrated a more pronounced drop. Scores in the AI-B-Hub group fell from approximately 6.0 to near 2.0, suggesting that personalized and adaptive music may be particularly effective at mitigating post-task stress in a busy environment. The NoAI groups also improved, but the reduction in stress was less dramatic, hovering near 3.2 in B-Hub and around 3.0 in Q-Lib. These results suggest that tailoring music to both environmental factors and self-reported stress may amplify the relaxation effect.



\begin{table}[h]
\centering
\caption{VAS-S Score Changing Pairwise Comparisons}
\begin{tabular}{lccccc}
\toprule
\textbf{Sample 1-Sample 2} & \textbf{Test Statistic} & \textbf{Std. Test Statistic} & \textbf{Sig.} & \textbf{Adj. Sig.\textsuperscript{a}} \\
\midrule
AI x B-Hub - NoAI x B-Hub       & 1.096 & 3.061  & 0.002  & 0.013 \\
AI x B-Hub - NoAI x Q-Lib       & 1.385 & 3.867  & <0.001 & 0.001 \\
AI x Q-Lib - NoAI x Q-Lib        & 1.250 & 3.491  & <0.001 & 0.003 \\
AI x Q-Lib - NoAI x B-Hub        & -0.962 & -2.685 & 0.007  & 0.043 \\
AI x B-Hub - AI x Q-Lib             & 0.135 & 0.376  & 0.707  & 1.000 \\
NoAI x B-Hub - NoAI x Q-Lib  & 0.288 & 0.806  & 0.420  & 1.000 \\

\bottomrule
\end{tabular}
\label{VASSWilcoxon}
\end{table}

Fig. \ref{VASChange} presents a box plot of \textit{VAS-S change scores} of four conditions (B-Hub/AI, B-Hub/NoAI, Q-Lib/AI, and Q-Lib/NoAI). \textit{VAS-S Change Score} represents the absolute difference between each participant’s VAS-S rating immediately after the first math task and after listening to the relaxing music. 
Table \ref{VASS Descriptive Statistics} summarizes the mean and standard deviation of overall VAS‐S changing scores.
The result suggests that participants in the two AI conditions display notably larger mean value decreases in stress compared to those in the NoAI conditions, suggesting that personalized music tended to relieve stress more effectively than pre-recorded tracks in both environments.


A repeated-measures ANOVA was conducted to examine the effects of the independent variables, Condition (AI vs. NoAI) and Environment (B-Hub vs. Q-Lib), on VAS-S changing scores over the four testing phases.
Mauchly's test indicated a violation of sphericity (\(\chi^2(5) = 13.523, \, p = .019\)), so the degrees of freedom were corrected using Greenhouse-Geisser (GG) estimates (\(\epsilon = 0.786\)). 
Results revealed a significant main effect of AI (AI vs. NoAI) on VAS-S, \( F(3, 23) = 12.135, \, p < .001 \), indicating that the AI led to a greater reduction in stress compared to the NoAI condition across all phases.
However, no significant main effect of Environment (B-Hub vs. Q-Lib) was found, suggesting that the testing environment did not significantly influence stress reduction.

A Wilcoxon signed-rank test was conducted for all six possible pairwise comparisons. As shown in Table~\ref{VASSWilcoxon}, the AI condition resulted in significantly lower stress scores than the NoAI condition in both busy (B-Hub) and quiet (Q-Lib) environments (\textit{p} < .05 after Bonferroni correction). In contrast, comparisons between environments (B-Hub vs. Q-Lib) within the same music type did not yield statistically significant differences (\textit{p} > .05), suggesting that stress reduction was primarily driven by the AI condition rather than environmental factors.

\subsection{Interview}

In this section, we analyze participants’ interview responses. We coded the user feedback under four emergent themes—reflecting core HCI concerns: \textbf{(1) Adaptability and Personalization}, \textbf{(2) User Engagement and Control}, \textbf{(3) Stress-Reduction Effectiveness}, and \textbf{(4) System Usability and Future Design Suggestions}. Representative quotations from participants illustrate each theme.

\subsubsection{Adaptability and Personalization}
A strong theme across interviews was the perceived adaptability of CAT music in addressing users’ stress levels. Most participants (22 of 26) mentioned the system’s ability to generate music suited to their \emph{current state} (stress level) and \emph{environment} (busy vs.\ quiet). This sense of personalization aligns with HCI principles that emphasize context-aware and user-centered design.

\begin{quote}
\textit{``The CAT music really resonated with me; it felt like it was created just for my stress level and environment, which helped me feel more at ease.''} (P4)
\end{quote}

Many participants contrasted CAT with YouTube music, noting that YouTube tracks felt ``static.'' CAT, in contrast, was described as \emph{evolving} with the user. Such feedback underscores the importance of dynamic adaptation in HCI systems.

\begin{quote}
\textit{``YouTube music is great, but it’s static. With CAT music, I felt like it adapted to my needs in a way that pre-recorded tracks couldn’t.''} (P5)
\end{quote}

\subsubsection{User Engagement and Control}
Beyond adaptation, participants also highlighted the \emph{interactive} nature of CAT music, which allowed them to influence the experience via \emph{prompts} and \emph{stress-level inputs}. This aspect boosted \emph{user engagement} by fostering a sense of \emph{control} and \emph{ownership} over the relaxation process—key concepts in HCI’s user-centered approach.

\begin{quote}
\textit{``I liked how the music responded to my stress level and the environment—it gave me a sense of control over my relaxation process.''} (P7)
\end{quote}

\begin{quote}
\textit{``It was refreshing to have music that felt like it was evolving with me rather than something preselected from a playlist.''} (P3)
\end{quote}

By enabling direct user involvement, CAT may have promoted deeper engagement—a recognized factor in interactive system success. Participants frequently framed this involvement as a \emph{collaborative} process with the system, suggesting the potential of \emph{co-creation} designs for stress-reduction tools.

\subsubsection{Stress-Reduction Effectiveness}
Most participants (24 of 26) reported a noticeable drop in subjective stress when using CAT, particularly in the \emph{quieter library setting}. Their remarks suggest that a \emph{tailored}, \emph{interactive} approach to music may enhance relaxation outcomes.

\begin{quote}
\textit{``In the quiet environment, the CAT music really helped me focus and relax; I could feel the tension leaving.''} (P11)
\end{quote}

\begin{quote}
\textit{``It felt more natural, like it belonged to the space I was in.''} (P8)
\end{quote}

This positive perception aligns with the \emph{quantitative findings} (VAS-S and GSR trends), where CAT consistently demonstrated stronger stress-reduction effects compared to standard YouTube music. In HCI terms, the context-aware adaptation likely contributed to higher \emph{perceived efficacy}.

\subsubsection{System Usability and Future Design Suggestions}
Participants also discussed practical aspects of using the CAT system, offering \emph{usability}-focused insights and \emph{feature requests}. They appreciated the straightforward interface and the ease of prompt selection:

\begin{quote}
\textit{``The interface is straightforward, and I like how easy it is to choose the prompts.''} (P1)
\end{quote}

However, several participants suggested adding more \emph{customization} features:

\begin{quote}
\textit{``It would be great if I could save my favorite prompts instead of selecting them every time.''} (P6)
\end{quote}

Additionally, participants highlighted the need for better \emph{real-time feedback}:

\begin{quote}
\textit{``Maybe the system could show how my stress levels are changing in real time, so I can adjust accordingly.''} (P4)
\end{quote}

Another common request was more \emph{diverse music options}:

\begin{quote}
\textit{``Having more genres to choose from would make the experience even better.''} (P10)
\end{quote}

Taken together, these suggestions provide valuable guidance for refining the system’s interface, personalization features, and content variety—key areas of focus in HCI to enhance user satisfaction and long-term engagement.

By categorizing feedback into these HCI-relevant themes, we gain insight into how an adaptive music system can be further optimized for stress reduction and user satisfaction. This analysis underscores the interplay between \emph{personalization}, \emph{engagement}, \emph{usability}, and \emph{effectiveness}—core tenets of HCI research—suggesting that future iterations of CAT should focus on refining these dimensions to enhance user experiences.

\section{Discussion}

The results of our user study indicate that personalized, Context-AI music (CAT) yields a more pronounced reduction in subjective stress than static, pre-recorded music from traditional relaxing music. In summary, participants reported lower VAS‐S scores when using CAT across both a quieter library (Q-Lib) and a busier hub (B-Hub), and these differences were statistically significant with large effect sizes. The significant interaction over time (Figs.\ref{VASlines} and \ref{VASChange}) shows that CAT music was especially effective in reducing stress after the math task and helped maintain lower stress levels until the end of the test. By contrast, the environment alone (Hub vs. Library) did not produce a significant difference in VAS‐S within each music condition, suggesting that the type of music intervention plays a pivotal role in overall stress reduction.

\textbf{Adaptive Music Personalization}
A key strength of CAT is its ability to tailor music according to user stress levels and environmental cues. Unlike fixed playlists, CAT modifies the generated music in real time, and participants often noted how this evolving experience helped them remain more attuned to their surroundings and internal stress. Still, achieving a balance between customization and predictability remains challenging: while users appreciated CAT’s adaptability, some wished for familiar elements or user-defined favorites. Integrating such familiarity could enhance comfort and continuity without undermining real-time adaptation. Future work may explore improved personalization algorithms that merge novelty, adaptability, and user familiarity, echoing broader HCI research on personalized music for stress management \cite{hamidi2019sensebox,ragone2020designing}.

\textbf{Context-Aware Stress Management}
Our findings also suggest that while CAT music reduced stress in both quiet and busy environments, the immediate impact appeared stronger in the calmer library setting, where fewer competing stimuli existed. This aligns with research suggesting that environmental distractions can moderate the efficacy of music-based interventions. Nonetheless, the lack of a statistically significant difference for environment alone implies that CAT’s adaptive features may offset some stressors in busier contexts. Integrating additional sensors (e.g., noise level or crowd density) could further refine context-awareness, automatically adjusting volume or musical complexity based on real-time feedback. Such sensor-enabled personalization might bolster CAT’s resilience across varied conditions.

\textbf{Expanded Applications Beyond Stress Relief}
Although we focused primarily on stress, participants and emerging literature both point to wider possibilities for adaptive music systems. Several users believed CAT could aid concentration during study or work sessions, consistent with work indicating that the right auditory environment can bolster productivity \cite{correa2009computer,lobo2019chimelight}. Moreover, adaptive music has shown promise in mental health and wellness domains, including anxiety management and sleep support \cite{grekow2021monophonic,hung2021emopia,Zhang_Zheng_Yang_Wei_Shan_Zhang_2024}.

Wearable integrations could deepen CAT’s capabilities by providing real-time physiological data (e.g., heart rate or skin conductance), enabling more precise musical adjustments. Collaborations with healthcare professionals or workplace wellness programs could then validate CAT’s applicability in wider contexts, ensuring it meets practical and clinical standards. Overall, participants consistently reported greater relief under the CAT condition, reflecting an HCI-centered approach where personalization and interactivity drive user engagement. Extending CAT with richer sensing and broader personalization options can evolve it into a versatile tool for well-being, productivity, and mental health support.

\section{Limitation}
As discussed in Section 3, the CAT system relies on the Suno API for music generation, limiting user control over the resulting tracks. Once a track is generated, it cannot be modified or refined to better suit user preferences without creating a new track each time. This restriction reduces the system’s flexibility in offering a more iterative and personalized experience. 

Another challenge lies in the scope of AI-generated music. Current models are trained on relatively small and stylistically constrained datasets, which may not fully address the varied musical tastes and stress relief needs of users. Expanding the diversity of available music styles could enhance the system’s capacity to meet a broader range of preferences.

Moreover, the system depends on self-reported stress levels and user-defined keywords, which can introduce subjective biases and inconsistencies. Integrating physiological sensors (e.g., heart rate variability or skin conductance) could bolster personalization accuracy by supplying real-time objective data and reducing the cognitive load on users.

Finally, while the system performed effectively in a controlled experimental environment, its utility in real-world contexts with unpredictable variables remains untested. Ongoing refinements and further research are necessary to improve the system’s adaptability and assure its sustained effectiveness in a variety of usage scenarios.
\section{Conclusion}
This paper introduced CAT, an AI-driven music generation system that personalizes relaxing music based on environmental cues and user-defined stress levels, addressing the limitations of traditional static playlists. By integrating visual inputs and self-reported stress indicators, CAT could interactively generate personalized music to reduce stress. 
Findings from our user study indicate that CAT significantly alleviates stress and enhances user engagement compared to pre-selected relaxing music. Despite these promising results, CAT has limitations, including the inability to refine generated tracks and the constrained variety of styles available in current AI models. Future work will focus on enhancing personalization, expanding musical diversity, and incorporating physiological sensors for more precise and dynamic adaptation. 

%
%
%
\bibliographystyle{splncs04}
\bibliography{custom}
%






\end{document}